\documentclass[showpacs,twocolumn,superscriptaddress,amsmath,amssymb,aps,prl]{revtex4}

\usepackage{graphicx}
\usepackage{dcolumn}
\usepackage{bm}
\usepackage{color}
\usepackage{xcolor}
\usepackage{amsfonts}
\usepackage{amssymb}
\usepackage{array}
\usepackage{times}
\usepackage{epsfig}
\usepackage{epstopdf}
\usepackage{setspace}
\usepackage{dcolumn}
\usepackage{amsmath}
\usepackage{latexsym}

\begin{document}

\title{On the spectral origin of non-Markovianity: an exact finite model}%

\author{Ruggero Vasile}
\author{Fernando Galve}
\author{Roberta Zambrini}
\affiliation{IFISC  (UIB-CSIC), Instituto de
F\'{\i}sica Interdisciplinar y Sistemas Complejos, UIB Campus, 07122 Palma
de Mallorca, Spain}
\date{\today}

\begin{abstract}

Non-homogeneous chain environments (e.g.  segmented ion traps)
are investigated through an exact diagonalization approach.
 Different spectral densities, including
band-gaps, can be engineered to separately assess memory effects.
Environment non-Markovianity is quantified with recently
introduced measures of information flow-back and non-divisibility of
the system dynamical map. By sweeping the bath spectrum via tuning
of the system frequency we show strongest memory effect at band-gap
edges and provide an interpretation
based on energy flow between system
and environment.
A  system weakly coupled to a stiff chain
ensures a Markovian dynamics, while the size of the environment as well as the
local density of modes
are not
substantial factors. We show an opposite effect when increasing the temperature
inside or outside the spectral band-gap. Further,
non-Markovianity arises for larger (negative and positive) powers of algebraic spectral densities,
being the Ohmic case not always the most Markovian one.

\end{abstract}


\maketitle
\par
\emph{Introduction}. Achievements in preparation, control and
measurements of quantum systems require a deep understanding of the
mechanism of interaction between a given  open  quantum system and
the surrounding environment \cite{WeiBreGar}. From a theoretical
point of view, popular approaches, e.g. derivations of master or
quantum Langevin equations \cite{WeiBreGar,Hanggi}, are based on the
assumption of an infinite heat bath with some given spectral density
$J(\omega)$ embedding all information about the
real couplings and frequencies in
the complex environment, and the structure of the coupling to the
system. Typical  approximations to simplify the treatment, such as
negligible memory effects (Markovian approximation), system time
coarse graining, weak system-bath coupling, large frequency cut-off,
Ohmic spectral density, drastically constrain the
possible frequency dependence of $J(\omega)$. Although simplified
spectral densities allow an analytical treatment, important
deviations from  these simple
instances  are common in several systems like, for instance,
electric circuits, acoustic polarons in metals and semi-conductors,
radiation damping of charged particles \cite{WeiBreGar}, photonic
crystals \cite{John,TanZha,PriPle,Hoeppe}, but also in ion traps
suffering electric field noise \cite{Hite}, micro-mechanical
oscillators \cite{aspelmeyerEisert} or polarized photons for broad
frequency spectra \cite{EXP_opt_NM}.

Memory effects and deviations from Markovian dynamics have been
widely explored considering the time dependence of master
equation-coefficients and  deviations from exponential decays
\cite{WeiBreGar,Haake,HPZ,nori} but only in the last few years
several approaches  have been proposed to systematically distinguish
Markovian evolution in terms of properties of the dynamical map
\cite{NM1,BLP,RHP,local_NM}. Different measures allow to quantify
non-Markovianity (NM) in terms of deviations from the Lindblad form
of the generator of the master equation \cite{NM1}, flow-back of
information from the environment \cite{BLP}, and entanglement decay
with an ancilla \cite{RHP}, to mention some of them. The recent
endeavor to better characterize memory effects in open systems not
only aims to a deeper understanding of dissipative dynamics in
physical, biological and chemical systems, but is also explored as a
resource in quantum technologies \cite{NMappl}.

The aim of this work is to identify non-Markovian effects
originating in the structure of the system and bath couplings as well as
in the distribution of energies, as given by the form of the
spectral density $J(\omega)$. To this end we consider a microscopic
model given by an inhomogeneous harmonic chain,
avoiding the limitations of approximate approaches. Finite models of
coupled oscillators have been used to assess entanglement dynamics
\cite{PleEis} and its generation when attaching ions to a chain
\cite{Morigi1} and can also provide an insightful test-bed  to
establish the regimes of validity of approximated master equations
\cite{RivPle}.
The case of an oscillator attached to a homogeneous chain was
already studied by Rubin to determine the statistical properties of
crystals with defects: this configuration leads to a Ohmic
dissipation (thus Markovian, at least for large temperature)
\cite{rubin}.
Moving to non-homogeneous chain configurations, we inquire the
origin of NM  to distinguish among several independent features
quantifying separately different sources of NM. When focusing on
periodic systems (e.g. dimers), we can engineer spectral densities with finite
band-gaps, like in semiconductors or photonic crystals
\cite{AshMer,John,TanZha,PriPle}.
 For suitable couplings we show that
the system is actually influenced by the resonant portion of the
environment. Memory effects \cite{BLP,RHP} are then evaluated by
sweeping the spectral density for a structured bath allowing us to
show the effects of the local form of the spectrum.


Before introducing the model, we point out that experimental
implementation of a tunable chain of oscillators can be obtained
through recent progress in segmented Paul traps
\cite{Kaler,Wineland} also allowing tunability of { ions }couplings and
onsite potentials. Other possible setups are based on photonic
crystal nanocavities, microtoroid resonators (see \cite{hartmann}
for a review) or optomechanic resonators \cite{aspelmeyerEXP}.
Furthermore, correlations spectra of the system can be measured to
gain insight on the spectral density induced by the rest of the
chain (see \cite{aspelmeyerEisert,nori}). The { non-homogeneous harmonic chain we consider in this work }
represents then a structured and controllable reservoir amenable to
experimental realization.

\par

\begin{figure}[h!]
\includegraphics[width=0.9\columnwidth]{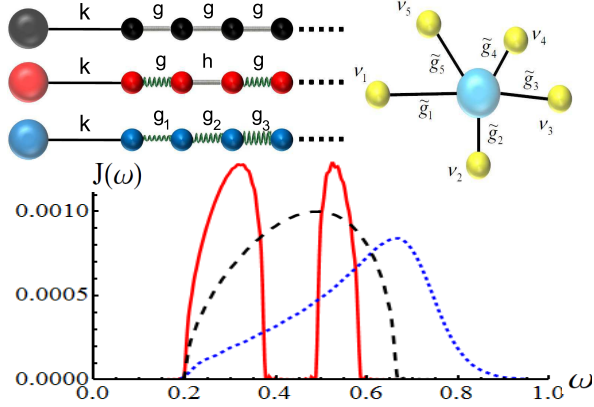}
\caption{(Colors online) Examples of spectral densities which can be
generated in our model: Rubin chain with $g=h=0.1$ and $k=0.01$
(black line), gapped spectral density with $g=0.1$, $h=0.05$ and
$k=0.0075$ (red line), and increasing coupling density with
$g_i=0.1+0.05\cdot i$ and $k=0.01$ (blue line). { Bath frequency $\omega_0=0.2$
(in unspecified, but fixed, frequency units) in all examples.
Schematic for the three plotted examples in the corresponding colors (top left)
and of a
star configuration (top right) also represented.} \label{Fig1}}
\end{figure}

\emph{The system: non-homogeneous chain}. We consider an open
quantum system consisting of an harmonic oscillator
$H_S=(p^2_S+\omega_S^2q^2_S)/2$, where $p_S$ and $q_S$ are the
system momentum and position operators, interacting with an
environment ($E$) that consists of an harmonic chain of $N$ elements
interacting through a spring-like coupling non-homogeneous along the
chain. A particularly dramatic deviation from the Ohmic spectrum
obtained with a homogeneous chain (Rubin model  \cite{rubin}) is
found when considering a periodic configuration (dimer) of identical
oscillators with alternate values of couplings $g$ and $h\leq g$
(see Fig. \ref{Fig1}) 
\begin{equation}
 H_E=\sum_i\bigl(p_i^2+\Omega_i^2q_i^2\bigl)/2-g\sum_{i}^{odd}q_iq_{i+1}
-h\sum_{i}^{even}q_iq_{i+1}
\end{equation}
with $\Omega_i=\sqrt{\omega_0^2+h+g}$ ($i=2...N-1$) and
$\Omega_1=\Omega_{N}=\sqrt{\omega_0^2+g}$ \cite{footnote1}. For any
$\omega_0$ and $g\neq h$ we build a band gap model, characterized by
a frequency spectrum distributed in an acoustic and
an optical band separated by a finite gap \cite{AshMer}.
%
The system is attached to the first element of the E chain and $k$
is a coupling constant with an interaction term of the form
$H_I=-kq_Sq_1$. We consider an initial factorized state
between system squeezed vacuum and a thermal bath at temperature
$T$. The chain configuration for real oscillators is mapped by
diagonalization of the environment (through an orthogonal
transformation $\mathbf{K}$) into a star configuration of
independent oscillators (bath's eigenmodes) of frequency $\nu_i$
interacting with the system with strength
$\tilde{g}_i=k\mathbf{K}_{1i}$ (see Fig. 1 and
\cite{SupMat}).

\emph{Generalized Langevin equation}. The reduced  dynamics of the system
is governed by a
Langevin equation, typically derived starting from a
star environment \cite{WeiBreGar,Hanggi},
\begin{equation}\label{Langevin}
\ddot{q}_S+\tilde{\omega}_s^2q_S+\int_0^t
ds\gamma(t-s)\dot{q}_S= {\xi}(t),
\end{equation}
where $ {\xi}(t)$ is Langevin forcing of the system
\cite{LangFor} and
$\gamma(t)=\sum_i\bigl(\tilde{g}_i^2/\nu_i^2\bigl)\cos\bigl(\nu_i
t\bigl)$ is the damping kernel accounting for dissipation and memory
effects.
The spectral density can be
obtained  from the damping kernel as
$J(\omega)=\omega\int_{0}^{\infty}\gamma(t)\cos(\omega t)dt$. For a
finite number $N$ of normal modes, the dissipative dynamics
suffers recurrence for times $\tau_R$ (related to reflection into
the system of the fastest signals traveling along the chain), whose value depends on
the number of modes $N$ and on the frequency spectrum \cite{Estim}.
If we look at times $t<\tau_R$ the spectral density can be described
by a smooth function of frequency. The form of this function can be
directly related to the dynamical properties of the system. For
instance, a linear --Ohmic-- spectral density , i.e.
$J(\omega)\propto\omega$, leads to 
a Markovian equation with time-independent friction kernel $-\gamma_0\dot{q}(t)$.
In the Rubin model it is possible to show that
$J(\omega)\propto\sqrt{\omega^2-\omega_0^2}\sqrt{\omega_c^2-\omega_S^2}\,\theta(\omega_c-\omega)$
with $\omega_c=\sqrt{2g}$ the highest (cutoff) frequency in the spectrum of
normal modes. Other non-homogeneous chain examples are shown in Fig.
\ref{Fig1}.

\emph{Resonance conditions}.
In order to investigate the role of the spectral density's shape, we first need
to know to what extent the system is affected by different bath eigenmodes
depending on their relative detuning. To do so we compare 
\begin{figure}[h!]
\begin{center}\includegraphics[width=0.99\columnwidth]{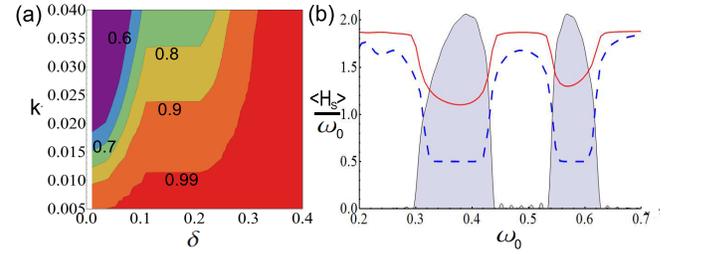}
\end{center}
\caption{(Colors online) (a) Averaged fidelity as a function of
system-bath coupling $k$ and bandwidth $\delta$, for a band gap
density with $g=0.1$, $h=0.05$ and $\omega_0=0.3$ at temperature
$T=0$. Integration performed up to $t_F=400$ for $N=50$ oscillators.
(b) System excitation number at $t=400$ as a function of the proper
frequency $\omega_S$ for a band gap model with  $g=0.1$, $h=0.05$
and $\omega_0=0.3$ ($J(\omega)$ as shaded area) at $T=0$.
System-bath coupling $k=0.005$ (red solid) and $k=0.025$ (blue
dashed).\label{Fig2}}
\end{figure}
the state of the system $\rho(t)$ with $\rho_{\delta}(t)$ obtained
by allowing the system to interact only with those normal modes
whose frequencies lie within a range $\delta$ to system frequency
$\omega_S$, i.e. such that $|\nu_i-\omega_S|<\delta$. The time
evolution is obtained by a full diagonalization of $H_S+H_R+H_I$
\cite{SupMat}. Deviation between
$\rho(t)$ and $\rho_{\delta}(t)$ can be witnessed by the Uhlmann
fidelity measure \cite{ZukBen},
$\mathbf{F}(\rho,\rho_{\delta})=Tr\sqrt{\sqrt{\rho}\rho_{\delta}\sqrt{\rho}}$.
This quantity averaged in time,
$\int_0^{t_F}\mathbf{F}(\rho,\rho_{\delta})dt/t_F$, is shown in Fig.
2a  for a band gap model with system frequency larger than the
optical band,
$\omega_S>\omega_c $. Here and in the following we choose
$t_F<\tau_R$ but big enough to explore the dissipative dynamics. If
we take the value $0.99$ as a guide for the eye, a weakly coupled
oscillator ($k<0.01$) interacts only with a small vicinity of modes
in the nearest band (right/optical band in Fig. \ref{Fig1}), while
already for $k=0.012$ we need to take into account modes in the
furthest band (left/acoustic band). This rough estimation allows to
appreciate that for weak couplings $k \ll\omega_c$ a reduced number
of resonant bath modes suffice to determine the system evolution.

The effect of the environment band-gap is clearly shown by looking
at the system excitation (average energy normalized by its
frequency) $\langle H_S(\tau_F)\rangle/\omega_S$ at $\tau_F$  by
varying the system frequency $\omega_S$ to sweep the spectral
density (Fig. \ref{Fig2}b). Energy is dissipated continuously when
$\omega_S$ is resonant with the bath while it can not propagate into
the chain for $\omega_S$ within the band-gap (leading to oscillatory
behavior and the formation of bound dressed states)
\cite{AshMer,John,TanZha,wang}.

\emph{Non-Markovian dynamics}. Among the different quantifiers of
NM appeared  recently in the literature
we consider hereafter the Breuer-Laine-Piilo (BLP)  \cite{BLP} and
Rivas-Huelga-Plenio (RHP)  \cite{RHP} measures. The first (BLP)
gives an interpretation of NM in terms of a back-flow of information
from the environment into the system. Its definition exploits the
contractivity property of the quantum trace distance $\mathbf{D}$
under completely positive and trace preserving
maps  \cite{BLP}.  
For continuous variable systems, and within Gaussian states, the definition has been
extended by substituting the trace distance with the fidelity
$\mathbf{F}$ \cite{VasGauNM,IncFid}, and the associated measure for
the degree of NM reads
\begin{equation}\label{NMBLP}
\mathbf{M}_{BLP}=\max_{\rho_1,\rho_2}\int_{dF/dt<0}\frac{dF(\rho_1,\rho_2)}{dt}dt
\end{equation}
where the maximization is taken among all pairs of Gaussian states
$(\rho_1,\rho_2)$ and integration is performed up to  $t_F$.

The RHP criterion witnesses the non-divisibility of the dynamical
map by preparing the system in an entangled state with an ancilla
and evaluating the non-monotonic time evolution of the entanglement.
The measure reads \cite{RHP}
\begin{equation}\label{NMRHP}
\mathbf{M}_{RHP}=\int_0^{\tau_R}\biggl|\frac{dE_{SA}}{dt}\biggl|dt-E_{SA}(\tau_R)+E_{SA}(0)
\end{equation}
where $E_{SA}(t)$ denotes a proper system-ancilla entanglement
measure, such as logarithmic negativity. We should stress that when
Eq. \eqref{NMRHP} gives zero, the map can be either divisible or
non-divisible. For details on the numerical evaluation of these measures see \cite{SupMat}.

\begin{figure}[h!]
\includegraphics[width=0.9\columnwidth]{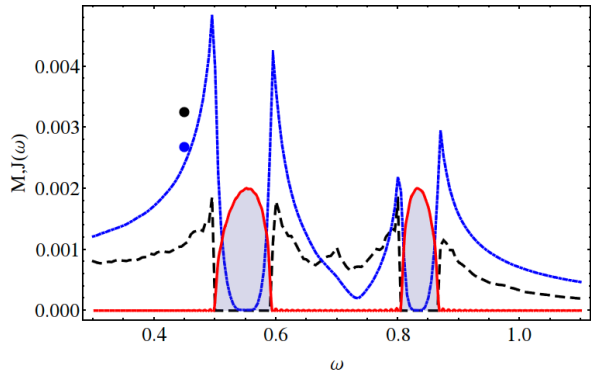}
\caption{(color online)
BLP non-Markovianity (blue) and RHP (black,dashed) for N=40 oscillators, $g=0.2$,
$h=0.05$, $k=0.001$, $\omega_0=0.5$ as a function of $\omega_S$, for
$T=0$. The red (continuous) curve
represents the spectral density used in arbitrary units. The extra dots are evaluated
at $\omega_s=0.45$ increasing the two-mode squeezing parameter from $r=1$ to $1.5$ (for RHP NM)
and increasing the squeezing parameter range for the BLP NM  maximization (from $r\in \{0.5,1\}$ to $r\in \{0.125,1.125\}$) \cite{SupMat}.
\label{Fig3}}
\end{figure}

In Fig. \ref{Fig3} we show both NM measures for a dimer chain
 environment  as a function of the
system frequency $\omega_S$ for temperature $T=0$. We find that local NM maxima
are present at the edges of the band-gap spectral density for both
measures, while rapidly decreasing both inside the band and the gaps
of the spectral density.
Sharp changes in frequency lead to long
times in the Fourier transform, and are responsible for broad noise
and dissipation kernels, for long time bath correlations
\cite{WeiBreGar,RivPle} and for pronounced NM signatures, Fig. \ref{Fig3}.
Full Markovian behavior is achieved only
inside the band where $\mathbf{M}_{BLP}\sim 0$. Even if the null
value of the RHP measure is inconclusive, we mention that increasing
the system-bath coupling $k$, it raises to finite values, yielding a
shape very much in agreement with the BLP measure. This is one the
few comparisons of two different NM measures in the literature
\cite{compare}.
We also notice  that in correspondence with the edges of the spectral
density $J(\omega)$, where  $\mathbf{M}_{BLP}$ is larger, we find a
higher density of modes. To exclude its connection with NM effects,
we artificially engineered the spectral density to obtain a constant
density of modes throughout the spectrum. This allows us to
establish that the enhancement of NM at the edges is actually  not
related to the normal mode density of states. The origin of NM is
elsewhere.

Interaction between system and environment ($H_I$)
leads to energy exchange. 
When the system is in resonance with a band of the spectral
density (optical or acoustical), energy is exchanged with many normal modes. In the real
chain picture this corresponds to energy allowed to travel along the
chain. This leads to an ever increasing indistinguishability of the states in Eq. \ref{NMBLP}
(all loosing energy and relaxing)
and thus to a low value of NM.
On the other hand, at the band edges and
in the band-gaps, the energy lost by the system cannot travel freely along the
chain, but bounces back and forth from the first elements of the chain to the system.
This implies a periodical increase/decrease of distinguishability of the states whose fidelity
we are integrating, hence we witness
a higher NM value. The higher the detuning (deeper in the band-gap), the less
excitations/energy are exchanged, resulting in a diminishing value for
the BLP NM.
This result is in accordance with Ref. \cite{nori}, where they study a generic bosonic/fermionic reservoir,
and show that band-gaps generate localized modes (thus dissipationless oscillatory behavior) plus
nonexponential decays (identified there with NM).
According to Ref. \cite{BLP},
Fig.  \ref{Fig3} provides a quantitative measure of NM at band-gap edges as flow back of information.

\emph{What makes a bath Markovian?} Besides the resonance conditions
and the back reflection of information/energy when the system is out
of resonance with the bath, what are the main factors leading to more
Markovian dynamics? We have tried to answer this question in a fivefold
approach, namely, with respect to the bath's size, its temperature, the width
of its spectral density, its shape (sub-Ohmic, Ohmic, super-Ohmic) and the
strength of its coupling to the system.

\begin{figure}
\begin{center}\includegraphics[width=\columnwidth]{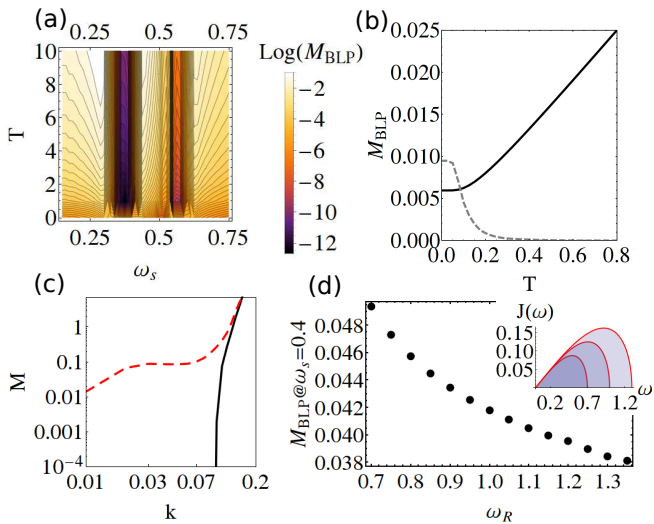}
\end{center}
\caption{
(color online) { (a)} log$(M_{BLP})$ for 50 oscillators, $g=0.1$, $h=0.05$, $k=0.001$ and $\omega_0=0.3$, for
temperatures $T\in[0,10]$ and $\omega_s$ (the limits of the acoustic and optical bands are obvious from
the discontinuity of $M_{BLP}$).
{ (b)} Temperature dependence of $M_{BLP}$ for $\omega_s=0.25~(0.375)$ in black (gray).
{ (c)} $M_{BLP}$ (dashed) and $M_{RHP}$ (continuous) at
$\omega_s=0.575$ for an homogeneous 50 oscillators chain, $g=h=0.1$ and $\omega_0=0.3$.
{ (d)} No-gap (star-configuration) spectrum $J(\omega)=k\omega\sqrt{\omega_R^2-\omega^2}/\omega_R$ with 40 oscillators, $k=0.00001$,
$\omega_s=0.4$ and varying $\omega_R$. (Inset shows $J(\omega)$ for three values of $\omega_R$.)
\label{Fig4}}
\end{figure}

It is known that at low temperatures, Markovian approximations are generally not valid
 \cite{WeiBreGar} and non-exponential decays of the correlation functions do arise \cite{Haake}.
In  Fig. \ref{Fig4}a we show that the back-flow  of information  $M_{BLP}$ is similarly present only at low temperatures ($T\lesssim\omega_c$) when
the system is dissipating within a band of the spectral density.
Surprisingly, when we move to a non-resonant
configuration, with the frequency of the system in the band-gap, NM is found to increase with temperature,
Fig. \ref{Fig4}b. When increasing temperature to higher values a linear increase of NM is
observed within the gap, while  in the band $M_{BLP}$ tends to vanish.

Most prominent NM effects are expected in the strong coupling regime
between system and bath \cite{WeiBreGar,Haake}: indeed reduction to
Markovian dynamics is provided by a decrease of the
 coupling, which reduces NM by two orders of magnitude by one order of magnitude decrease in coupling.
For the $M_{RHP}$ we see that it tends to zero very fast (remember
that a zero value is inconclusive) Fig. \ref{Fig4}c. The importance
of frequency cut-off in the spectral density is relevant not only to
avoid unphysical divergences but also when discussing NM in terms of
relevant time scales. Interestingly, a chain environment allows to
engineer the value of the frequency cut-off of an Ohmic environment
by increasing the coupling strengths within the chain. As shown in
Fig. \ref{Fig4}d an increase of the width of the spectrum (stiff
chains) allows for a
monotonic flow of information into the environment leading to a
decrease of NM. We also checked the dependence of NM with the bath
size, seeing none (except for the role of the recurrence time).

\begin{figure}[h!]
 \includegraphics[width=\columnwidth]{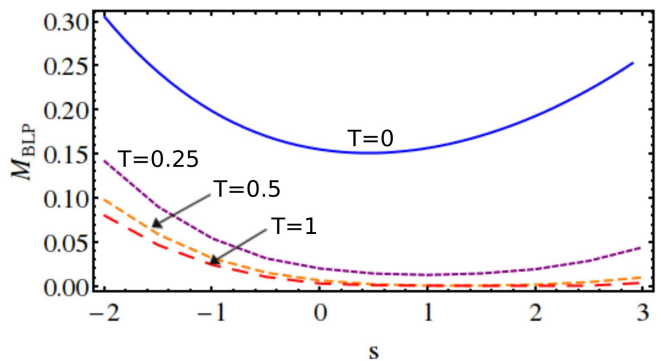}
\caption{
$M_{BLP}$ for the star-configuration density $J(\omega)=k(\omega/\omega_S)^s\Theta(\omega-0.25)\Theta(0.75-\omega)$ ($\Theta$ is the step function) as a function of $s$ at $\omega_s=0.5$, for temperatures
$T=0,0.25,0.5,1$.
\label{Fig5}}
\end{figure}

Finally we  built an artificial star model (Fig. \ref{Fig1}) with equally spaced
frequencies and couplings $\tilde{g}_i$ in order to isolate the
effects of different  algebraic behavior of the spectrum, looking,
as common in the literature, at deviations from the Ohmic case. In
Fig. \ref{Fig5} we show $M_{BLP}$ for algebraic spectral densities
$J(\omega)\sim  \omega^s $
 with fixed sharp cutoffs at $0.25,0.75$. We consider both
positive and negative algebraic forms  \cite{aspelmeyerEisert} and
observe an increase of NM in both cases when departing from $s_{\rm
min}\sim 1/2,1,1.4,2,2.8$ at temperatures $T=0,0.25,0.5,1,2$ ($T=2$
not shown).

Surprisingly, the lowest NM is achieved for $s=1/2$ at zero temperature, instead of $s=1$ (Ohmic case). For higher temperatures
the lowest value is achieved for $s=1$ or a bit higher. Also, a bath with $J(\omega)\sim 1/\omega^2$ is more non-Markovian than one with $J(\omega)\sim \omega^2$.
 We also stress that different scaling and constraints  of power functions  densities lead to
significant influence on memory effects \cite{SupMat}.

\emph{Summary and conclusions.}

We have considered different (non-phenomenological) spectral
densities attainable by tuning non-homogeneous oscillators chains
that can be implemented in segmented ion chains traps and also with
nano-oscillators. A system attached at one extreme of the chain
dissipates in this finite bath and exhibits memory effects.
 The importance of our analysis is
the separate assessment, without approximations,  of environment features that can be microscopically
engineered and the quantitative comparison of
two NM measures \cite{BLP,RHP} to establish their influence in retaining environment memory effects.

The possibility to tune the chain in a dimer configuration
allows us to assess the influence of band-gaps and  to obtain a more detailed picture on the origin of
Markovianity in relation to specific features of the spectral
density. The main role describing the behavior of NM is played by
a resonance condition: if the system is resonant with the
normal modes of the bath, energy transfer along the chain is allowed
and therefore information and energy flow irreversibly from system
to bath leading to a Markovian dynamics.
The largest flow back of information is found
at the edges of the gaps  where the energy bounces between the system
and bath.

A high frequency cutoff and weak coupling, the first
obtained by increasing the stiffness of the chain and the latter by
decreasing the system-bath coupling, are major factors for ensuring
Markovian dynamics.  Indeed, NM effect of strong coupling or frequency
cut-off, have already been discussed in the literature and we find consistent results.
A relevant point when dealing with finite systems is
that actually neither the size of the bath is important
(only matters in limiting the recurrence time) nor the
local density of modes has any significant influence on memory effects.

On the other hand, unexpected results have been found when increasing temperature
leading to opposite behavior inside and outside the band with
a (linear in T) NM increase in the band-gap while inside the band memory
effects become negligible for temperatures larger than the
environment frequencies.
Finally we have  quantified NM when departing from the Ohmic case,
considering positive and negative algebraic spectral densities  $J(\omega)\propto
\omega^s$ for $-2\le s \le 3$  (being $s=-2$ a recently measured value \cite{aspelmeyerEisert}) showing an enhancement of memory effects for both positive and negative
values of $s$,  with the Ohmic case (s=1) not being necessarily the most Markovian one.

\begin{acknowledgments}

Financial support from MICINN, MINECO, CSIC, EU commission and FEDER funding under grants
FIS2007-60327 (FISICOS), FIS2011-23526 (TIQS), post-doctoral JAE program and COST Action MP1209
is acknowledged.

\end{acknowledgments}

\section{Supplementary information}

 \subsection*{Environmental diagonalization}

 Let's consider the Hamiltonian $H_E$ and rewrite it in the
 following way
 \begin{equation}\label{SMHE}
 H_E=\frac{\mathbf{p}^T\mathbf{p}}{2}+\mathbf{q}^T\mathbf{A}\mathbf{q}
 \end{equation}
 where we compact the position and momentum operators in a vector
 formalism, i.e. $\mathbf{q}\equiv\{q_1,q_2...,q_N\}^T$ and
 $\mathbf{p}\equiv\{p_1,p_2...,p_N\}^T$. Moreover we have an $N\times
 N$ matrix $\mathbf{A}$ with elements
 $\mathbf{A}_{ij}=\Omega_i^2\delta_{ij}/2-\mathbf{G}_{ij}/2$, where
 the connection matrix $\mathbf{G}$ has the following form
 \begin{equation}\label{GMat}
 \mathbf{G}=\left(
          \begin{array}{cccccc}
        0 & g & 0 & 0 & ... & 0 \\
        g & 0 & h & 0 & ... & 0 \\
        0 & h & 0 & g & ... & 0 \\
        0 & 0 & g & ... & ... & ... \\
        ... & ... & ... & ... & 0 & g \\
        0 & 0 & 0 & ... & g & 0 \\
          \end{array}
        \right)
 \end{equation}
 Since $\mathbf{A}$ is symmetric, it can be diagonalized by an
 orthogonal transformation $\mathbf{K}$, i.e.
 $\mathbf{K}^T\mathbf{A}\mathbf{K}=\mathbf{D}$ where $\mathbf{D}$ is
 a the diagonal matrix containing the eigenvalues $\lambda_i$ of
 $\mathbf{A}$. Thus defining the new variables
 $\mathbf{Q}=\mathbf{K}^T\mathbf{q}$ and
 $\mathbf{P}=\mathbf{K}^T\mathbf{p}$, we can rewrite the hamiltonian
 as
 \begin{equation}\label{SMHE2}
 H_E=\sum_{i=1}^N\biggl[\frac{P_i^2}{2}+\frac{\nu_i^2Q_i^2}{2}\biggl]
 \end{equation}
 where the eigenfrequencies are $\nu_i=\sqrt{2\lambda_i}$,and again
 $\mathbf{Q}\equiv\{Q_1,Q_2...,Q_N\}^T$ and
 $\mathbf{P}\equiv\{P_1,P_2...,P_N\}^T$. Thus we passed from a chain
 environment into the equivalent star model.

 \subsection*{Full Diagonalization and time evolution}

 The starting point is the total Hamiltonian $H=H_S+H_E+H_I$ after
 the diagonalization of the environment. Since $H$ is also quadratic
 in position and momentum operators, it can again be written as
 \begin{equation}\label{SMH}
 H=\frac{\mathbf{p}^T\mathbf{p}}{2}+\mathbf{q}^T\mathbf{B}\mathbf{q}
 \end{equation}
 where, contrary to the notation of last section, we have
 $\mathbf{q}\equiv\{q_1,q_2...,q_N,q_S\}^T$ and
 $\mathbf{p}\equiv\{p_1,p_2...,p_N,p_S\}^T$. The $(N+1)\times(N+1)$
 matrix $\mathbf{B}$ has elements $\mathbf{B}_{ii}=\nu_i^2/2$ for
 $i=1....N$, $\mathbf{B}_{N+1N+1}=\omega_S^2/2$ and
 $\mathbf{B}_{i,N+1}=\mathbf{B}_{N+1,i}=-\tilde{g}_i/2$.

 We can again perform a diagonalization of $\mathbf{B}$ through an
 orthogonal matrix $\mathbf{O}$, i.e.
 $\mathbf{O}^T\mathbf{B}\mathbf{O}=\mathbf{F}$, and upon defining the
 new system-environment normal modes
 $\mathbf{Q}=\mathbf{O}^T\mathbf{q}$ and
 $\mathbf{P}=\mathbf{O}^T\mathbf{p}$, we rewrite the full Hamiltonian
 as
 \begin{equation}\label{SMH2}
 H=\sum_{i=1}^{N+1}\biggl[\frac{P_i^2}{2}+\frac{f_i^2Q_i^2}{2}\biggl]
 \end{equation}
 where $\sqrt{2 f_i}$ are the eigenvalues of $\mathbf{B}$ contained
 in the diagonal matrix $\mathbf{F}$.

 In this picture, the time evolution for each normal mode is trivial,
 \begin{equation}\begin{split}\label{TimeEvo1}
 &Q_i(t)=Q_i(0)\cos\bigl(f_it\bigl)+\frac{P_i(0)}{f_i}\sin\bigl(f_it\bigl)\\
 &P_i(t)=-f_iQ_i(0)\sin\bigl(f_i t\bigl)+P_i(0)\cos\bigl(f_i t\bigl)
 \end{split}\end{equation}
 Now, remembering the new variables are connected to the old ones
 through the orthogonal transformation $\mathbf{O}$ at any time $t$,
 we easily get
 \begin{equation}\begin{split}\label{TimeEvo2}
 &q_i(t)=\sum_{j=1}^{N+1}\bigl[\mathbf{B}^{QQ}_{ij}(t)q_j(0)+\mathbf{B}^{QP}_{ij}(t)p_j(0)\bigl]\\
 &p_i(t)=\sum_{j=1}^{N+1}\bigl[\mathbf{B}^{PQ}_{ij}(t)q_j(0)+\mathbf{B}^{PP}_{ij}(t)p_j(0)\bigl]
 \end{split}\end{equation}
 where
 \begin{equation}\begin{split}\label{DefMatEvo}
 &\mathbf{B}^{QQ}(t)=\mathbf{O}\cdot\mathbf{Co}\cdot\mathbf{O}^T\\
 &\mathbf{B}^{QP}(t)=\mathbf{O}\cdot\mathbf{\frac{Si}{f}}\cdot\mathbf{C}^T\\
 &\mathbf{B}^{PQ}(t)=-\mathbf{O}\cdot f\mathbf{Si}\cdot\mathbf{O}^T\\
 &\mathbf{B}^{PP}(t)=\mathbf{O}\cdot\mathbf{Co}\cdot\mathbf{O}^T
 \end{split}\end{equation}
 where $\mathbf{Co}$ is a diagonal matrix with
 $\mathbf{Co}_{ii}=\cos(f_it)$, $f\mathbf{Si}$ is also diagonal such
 that $f\mathbf{Si}_{ii}=f_i\sin(f_it)$ and
 $\frac{\mathbf{Si}}{f}_{ii}=\sin(f_it)/f_i$. Eqs. \eqref{TimeEvo2}
 provide the time evolution of the position and momentum operators
 for the system and the normal modes when their form at time zero is
 given.

\subsection*{Numerical evaluation of NM measures}

The simulations for $M_{RHP}$ have been done  for two-mode vacuum
squeezed states between system and ancilla, with a squeezing
parameter $r=1$ unless otherwise stated. As shown in Fig.3 in the
main paper (extra black dot), increasing this parameter yields
higher values for the measure. However, there are not qualitative
changes in the  behavior of $M_{RHP}$ (see Fig.~\ref{Figapp2}).
\begin{figure}
\includegraphics[width=0.85\columnwidth]{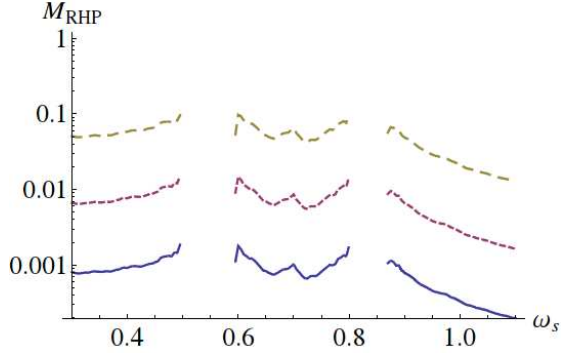}
\caption{ RHP non-Markovianity against system frequency $\omega_s$
comparing the result with $r=1,2,3$ (continuous, dashed,
dashed-dashed). The behavior is the same, only with higher absolute
values. \label{Figapp1}}
\end{figure}

The main issue with this measure is that for low system-bath
couplings it is inconclusive (i.e. $M _{RHP}=0$), but this can be
easily fixed by increasing $k$. Typical time steps for integration
of this measure $\delta t=0.1$ (e.g. for $\tau_R\sim 700$ in Fig.3)
have been used, which is more than enough to resolve the positive
slopes of the dynamics for $E_{SA}$ (see eq. (4) in main work).

\begin{widetext}

\begin{figure*}[h!]
\includegraphics[width=\columnwidth]{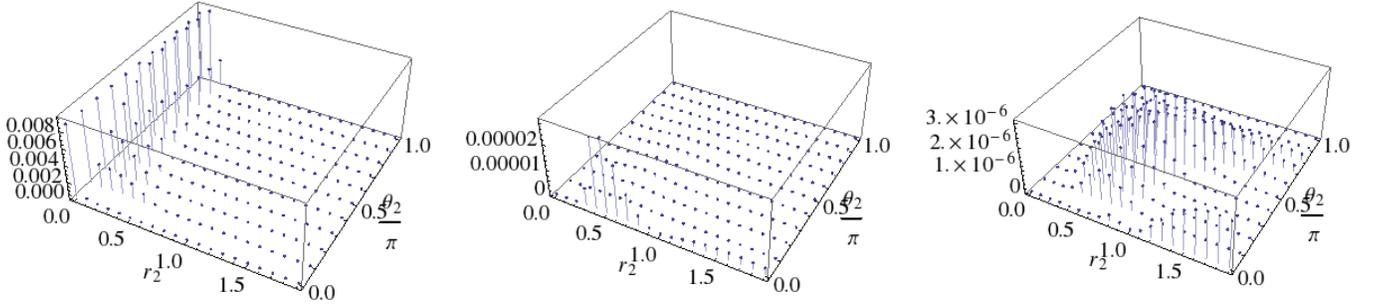}
\caption{ BLP non-Markovianity as used for the figures related to
temperature, for $T=0,0.5,1$ and $\omega_s=0.375$. It can be seen
that for higher temperatures several maxima appear at different
squeezing parameters $r_2,\theta_2$. \label{Figapp2}}
\end{figure*}

\end{widetext}

The simulations for $M_{BLP}$ have been performed between pairs of
one-mode vacuum squeezed states. In Fig.3 of main work, one of them
having squeezing parameter $r_1=1$ and phase-space angle
$\theta_1=0$, the other state having $r_2\in[0.5,1]$, $\Delta
r_2=0.5$ and $\theta_2\in[0,\pi/2]$ with $\Delta\theta_2=\pi/4$
(same for Fig. 4c; for Figs. 4d and 5 we have used $r_2\in[0.25,1]$
with $\Delta r_2=0.25$). The bold (blue) point in Fig. 3 in the main
paper was obtained by increasing $r_2\in[0.125,1.125]$, $\Delta
r_2=0.5$ and $\theta_2\in[0,\pi/2]$ with $\Delta\theta_2=\pi/4$.
Instead, for figures 4a,b we have used a much more thorough scan
with $r_1=1/3,1$, $\theta_1=0$, $r_2\in[0,2]$, $\Delta r_2=0.1$ and
$\theta_2\in[0,\pi]$ with $\Delta\theta_2=\pi/10$. We stress that
the behavior against temperature is quite sensitive to the range of
squeezings used and needs to be scanned intensively, because the
optimal pair for this measure depends on the bath temperature
(mostly for low T). We show an example in Fig.~\ref{Figapp2}.

\subsection*{NM behavior with non-Ohmic bath for other
parametrizations}

\begin{figure}
\includegraphics[width=0.85\columnwidth]{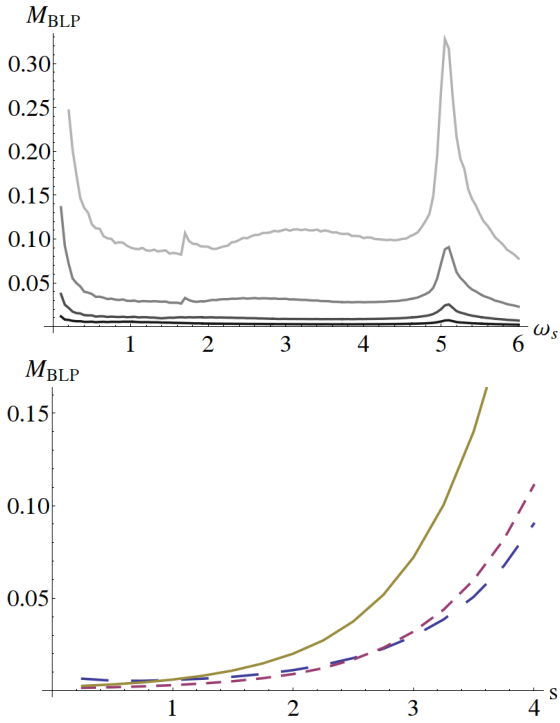}
\caption{ Top: BLP non-Markovianity against system frequency
$\omega_s$ with
$J(\omega)=k(\omega-\omega_0)^s\Theta(\omega_R-\omega)$,
$\omega_0=1$, $\omega_R=5$ and $s=1,2,3,4$ (black for $s=1$ to gray
for $s=4$. Bottom: BLP NM versus $s$ evaluated at $\omega_s=1,3,5$
(sparse dashed, tight dashed, continuous) for the spectral density
above. \label{Figapp3}}
\end{figure}

The behavior of $M_{BLP}$ under the non-Ohmic spectral density
\begin{equation}
J(\omega)=k(\omega/\omega_s)^s\Theta(\omega-0.25)\Theta(0.75-\omega)
\end{equation}
 was shown in last figure of main work,
Fig.  5. This is a spectral density which pivots around $\omega_s$
and therefore keeps the coupling strength ($J(\omega_S)=k\ ,\forall
s$) fixed, while at the same time the extreme points ($\omega_0$ and
$\omega_R$) it differs in values. However, choosing for example
\begin{equation}
 J(\omega)=k(\omega-\omega_0)^s\Theta(\omega_R-\omega)
\end{equation}
(which keeps $J(\omega_0)$ constant but differs at $J(\omega_R)$ for
different $s$) would change the value of $M_{BLP}$, as shown in Fig.
\ref{Figapp3}.

\begin{figure}
\includegraphics[width=\columnwidth]{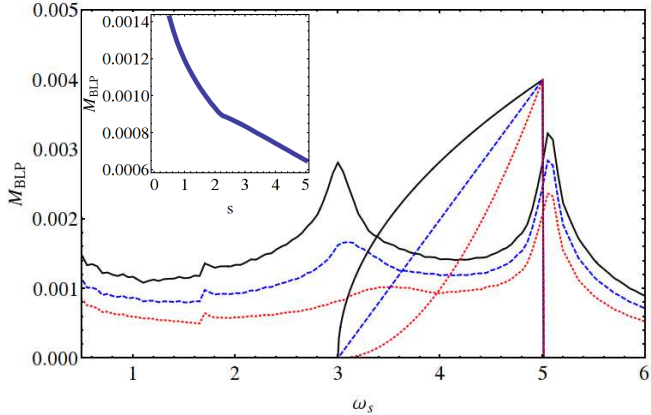}
\caption{ BLP non-Markovianity against system frequency $\omega_s$
with $J(\omega)=k[(\omega-\omega_0)/(\omega_R-\omega_0)]^s
\Theta(\omega_R-\omega)$, $\omega_0=3$, $\omega_R=5$ and $s=1/2,1,2$
(continuous, dashed, dotted). We have drawn the spectral density to
guide the eye. Inset: BLP NM versus $s$ evaluated at $\omega_s=4$.
\label{Figapp4}}
\end{figure}

Deeper differences are found considering different power law spectra
densities with constraint coupling strengths at both $\omega_0$ and
$\omega_R$, as results from
\begin{equation}
 J(\omega)=k[(\omega-\omega_0)/(\omega_R-\omega_0)]^s\Theta(\omega_R-\omega).
\end{equation}
In Fig. \ref{Figapp4} it can be seen that the behavior of $M_{BLP}$
is again different. It does  actually decrease in the super-ohmic
case (larger $s$ values).

We notice that in the latter two cases the spectral density  at the
system frequency $J(\omega_S)$ differs in value when changing  $s$,
unlike the case discussed in the main work. These examples show that
normalization has to be carefully taken into account when drawing
general conclusions about non-Markovian aspects of dissipation in
presence of different power law in the spectral density.


\begin{thebibliography}{10}
\bibitem{footnote1} Translation invariance of the environment is recovered by imposing an extra local potential
to the first and last masses. When $g=h$ and $\omega_0=0$, in the
limit $N\rightarrow\infty$ a Rubin chain is recovered \cite{rubin}.
\bibitem{Estim}
A rough estimate of the recurrence time can be given as the time it
takes for the fastest phonon to travel back and forth along the
chain. The group velocity of each phonon is simply $v_k=d\nu_k/dk$.
\bibitem{LangFor}
For the sake of completeness we provide the expression of the
Langevin force ${\xi}(t)=\sum_i\tilde{g}_i\bigl(q_i\cos(\nu_i
t)+p_i/\nu_i\sin(\nu_i t)\bigl)-\gamma(t)q_S(0)$.
\end{thebibliography}
\end{document}